\documentclass{PoS}
\usepackage{epsf,amsmath,bbold,amsfonts,stmaryrd,amssymb,mathrsfs}
\usepackage{url}
\usepackage{graphicx}
\usepackage{tikz}
\usetikzlibrary{arrows,matrix,positioning}
\usepackage{dsfont}
\usepackage{bm}
\usepackage{fontenc}
\usepackage{times}
\usepackage{mathptmx}
\usepackage[numbers,square,comma,compress,merge]{natbib}

\def\be{\begin{equation}}
\def\ee{\end{equation}}

\def\bea{\begin{eqnarray}}
\def\eea{\end{eqnarray}}

\def\ba{\begin{array}}
\def\ea{\end{array}}

\def\bc{\begin{center}}
\def\ec{\end{center}}

\def\a{\alpha}

\def\g{\gamma}

\def\d{\delta}

\def\m{\mu}

\def\x{\xi}

\def\op{\mathcal O}

\def\f{\frac}
\def\l{\left}

\def\p{\partial}
\def\r{\right}
\def\Tr{\text{tr}}

\newcommand{\av}[1]{\langle #1\rangle}

\makeatletter

\newsavebox\myboxA
\newsavebox\myboxB
\newlength\mylenA

\newcommand*\xoverline[2][0.75]{%
\sbox{\myboxA}{$\m@th#2$}%
\setbox\myboxB\null
\ht\myboxB=\ht\myboxA%
\dp\myboxB=\dp\myboxA%
\wd\myboxB=#1\wd\myboxA
\sbox\myboxB{$\m@th\overline{\copy\myboxB}$}
\setlength\mylenA{\the\wd\myboxA}
\addtolength\mylenA{-\the\wd\myboxB}%
\ifdim\wd\myboxB<\wd\myboxA%
\rlap{\hskip 0.5\mylenA\usebox\myboxB}{\usebox\myboxA}%
\else
\hskip -0.5\mylenA\rlap{\usebox\myboxA}{\hskip 0.5\mylenA\usebox\myboxB}%
\fi}
\makeatother

\title{Aspects of spontaneous breaking of conformal invariance in the
Fishnet CFT}

\ShortTitle{Spontaneous conformal symmetry breaking in Fishnet CFT}

\author{\speaker{Georgios K. Karananas}\\ Arnold Sommerfeld Center,
        Ludwig-Maximilians-Universit\"at M\"unchen,\\
        Theresienstra{\ss}e 37, 80333, M\"unchen, Germany\\
        E-mail: \email{georgios.karananas@physik.uni-muenchen.de}}

\abstract{I will discuss spontaneous conformal symmetry breaking in
the strongly $\gamma$-deformed limit of the $\mathcal N=4$
supersymmetric Yang-Mills theory known as~\emph{Fishnet Conformal
Field Theory}.}

\FullConference{Corfu Summer Institute 2019 "School and Workshops on
		Elementary Particle Physics and Gravity" (CORFU2019)\\
		31 August - 25 September 2019\\ Corfu, Greece}

\begin{document}

\section{Intoduction and motivation}

This talk\,\footnote{The slides can be found online at:\\
\texttt{\scriptsize{
http://physics.ntua.gr/corfu2019/Talks/georgios\_karananas@physik\_uni-muenchen\_de\_01.pdf}}~.}~
is based on the findings of~ \cite{Karananas:2019fox} written in
collaboration with Vladimir Kazakov and Mikhail Shaposhnikov, and
concerns the spontaneous breakdown of conformal symmetry in a specific
limiting case of the $\gamma$-deformed $\mathcal N=4$~supersymmetric
Yang-Mills (SYM) theory, the~\emph{Fishnet CFT}~(FCFT).
    
The model was discovered a few years ago by Gurdogan and
Kazakov~\cite{Gurdogan:2015csr} and since then it is being extensively
studied~\cite{Gromov:2017cja,Chicherin:2017frs,
Grabner:2017pgm,Kazakov:2018qez,Gromov:2018hut,Basso:2017jwq,
Basso:2018agi,Derkachov:2018rot,Korchemsky:2018hnb,Ipsen:2018fmu,
Basso:2018cvy,Kazakov:2018gcy,deMelloKoch:2019ywq,Gromov:2019aku,
Gromov:2019bsj,Chowdhury:2019hns,Gromov:2019jfh,Adamo:2019lor,
Basso:2019xay,Alfimov:2020obh}, although the majority of the
considerations concern the phase where the conformal symmetry is not
broken. The reasons the FCFT has attracted considerable attention are
manyfold. Among others, in the planar $N_c\to \infty$ limit, although
nonsupersymmetric, the theory is a genuine CFT, it appears to be
integrable~\cite{Gurdogan:2015csr,Caetano:2016ydc,Gromov:2017cja} and
in addition it accommodates a rich set of flat directions
corresponding to spontaneously broken quantum conformal
symmetry~\cite{Karananas:2019fox}.
  
Contrary to what happens if the symmetry is unbroken, on top of
nontrivial (i.e. symmetry-breaking) flat vacua, the spectrum of the
theory---apart from a massless dilaton---comprises massive particle
states. Nevertheless, and in spite of the presence of mass
scale(s), the vacuum energy is exactly zero (for instance
see~\cite{Amit:1984ri,Einhorn:1985wp,Rabinovici:1987tf,
Shaposhnikov:2008xi}). This is due to the stringent constraints that
conformal invariance imposes on a system. Specifically, it dictates
that the potential be a homogeneous function of the fields. In turn,
if there exist nontrivial configurations that extremize the potential,
then automatically the vacuum energy of the theory vanishes. In the
FCFT this happens naturally, i.e. without resorting to finetunings. In
general, such solutions were believed to be present only at specific
values of the corresponding coupling(s) for nonsupersymmetric
theories.

Apart from a possible ``academic'' interest on the moduli space of the
FCFT---especially in connection with the parent $\mathcal N = 4$~SYM
and its $\gamma$-deformation---the existence of natural flat
directions which are not lifted by quantum corrections can be of
relevance in particle physics phenomenology too. It is conceivable
that some of the features of the FCFT concerning the natural breakdown
of quantum scale/conformal symmetry may be universal and be also
present in more realistic theories.\footnote{For a non-exhaustive list
of relevant works
see~\cite{Englert:1976ep,Wetterich:1987fk,Wetterich:1987fm,
Wetterich:1994bg,Bardeen:1995kv,Meissner:2006zh,Shaposhnikov:2008xb,
Shaposhnikov:2008xi,Shaposhnikov:2009nk,Blas:2011ac,GarciaBellido:2011de,
Bezrukov:2012hx,Armillis:2013wya,Tavares:2013dga,Gretsch:2013ooa,
Khoze:2013uia,Rubio:2014wta,Boels:2015pta,Karam:2015jta,Karananas:2016grc,
Karananas:2016kyt,Karam:2016rsz,Ferreira:2016wem,Rubio:2017gty,
Casas:2017wjh,Ferreira:2018itt,Gorbunov:2018llf,Casas:2018fum,
Mooij:2018hew,Shaposhnikov:2018nnm,Karam:2018jcz,Ferreira:2019zzx,
Ghilencea:2019rqj,Ghilencea:2020piz} and references therein, as well
as~\cite{Wetterich:2019qzx} for a recent nice
review. \label{foot:refs} }~This may in turn put the cosmological
constant problem in a different context. Note that the latter theories
may provide the appropriate language to address yet another
fine-tuning puzzle of the Standard Model: the gauge hierarchy
problem. It has been argued that the smallness of the Higgs mass as
compared to the Planck scale may be the consequence of scale/conformal
symmetry~\cite{Wetterich:1983bi, Bardeen:1995kv} in combination with
no particle thresholds between the electroweak and Planck
scales~\cite{Shaposhnikov:2007nj,Shaposhnikov:2008xi},\footnote{Interestingly,
the absence of heavy particle states may also be achieved in the
context of Grand Unified Theories, see~\cite{Karananas:2017mxm}.}~and
the presence of nonperturbative gravitational effects operative at
very high
energies~\cite{Shaposhnikov:2018xkv,Shaposhnikov:2018jag,Shkerin:2019mmu,
Shaposhnikov:2020geh}.

The paper is organized as follows. A (very) basic overview of the FCFT
is given in Sec.~\ref{sec:FCFT}. Classical and quantum aspects of the
symmetry breaking vacua are discussed in
Sec.~\ref{sec:flat_directions}. The conclusions are presented in
Sec.~\ref{sec:conclusions}.

\section{A crash course on the basics of Fishnet CFT}
\label{sec:FCFT}

The starting point of the discussion is the purely scalar sector of
the~$\gamma$-deformed $\mathcal N=4$~SYM Lagrangian (for example
cf.~\cite{Fokken:2013aea,Gurdogan:2015csr})\,\footnote{In what
follows, all considerations concern four spacetime dimensions
exclusively. The generalization to arbitrary number of dimensions was
considered in~\cite{Kazakov:2018qez}.}
\be
\label{eq:gam_def}
\mathscr L = N_c\,\Tr \l [ \p_\m \xoverline \phi _ a \p _ \m \phi _ a
+ g_{YM}^2 \l(\f 1 4 \{\xoverline \phi _ a ,\phi _ a \} \{\xoverline
\phi _ b ,\phi _ b \} -e^ {-i\epsilon_{abc}\g_c}\xoverline \phi_a
\xoverline \phi_b \phi _a \phi _b \r)\r] \ ,
\ee
where the $\phi_a$'s are 3 complex scalar (traceless) $N_c\times N_c$
matrices in the adjoint of $SU(N_c)$, a bar denotes Hermitian
conjugation, $g_{YM}$ is the Yang-Mills coupling, and the $\g_a$'s are
the parameters of the deformation, also called ``twists.''~In the
above, summation over all repeated spacetime $(\mu=0,\ldots,3)$ as
well as internal $(a,b,c=1,2,3)$ indexes is tacitly assumed; as
customary, the curly brackets stand for the anticommutator and
$\epsilon_{abc}$ is the three-dimensional totally antisymmetric
symbol.  The $\g$-deformed theory is defined by~(\ref{eq:gam_def}),
supplemented by the corresponding gauge and fermionic sectors, which
are not explicitly written down since they play no role in the
subsequent considerations.

To obtain the FCFT one takes the double-scaling (DS) limit of weak
Yang-Mills coupling, while keeping $\g_{1,2}$ fixed and forcing 
$\g_3$ to be large and imaginary
\be
\label{eq:DS_limit}
g_{YM} \to 0 \ ,
~~~\g_{1,2}\to\text{fixed}\ ,~~~
\g _ 3 \to + i \times \infty\ ,
\ee
such that
\be
\label{eq:DS_limit_2}
\xi^ 2 _ {1,2} = g_{YM} ^ 2 \,N_ c e^{-i\g_{1,2}} \to 0 \
,~~~\xi ^ 2 =g_{YM} ^ 2 \,N_ c e^{-i\g_3}\neq 0 \ .
\ee
One can show that the gauge fields, fermions and the scalar $\phi_3$
decouple~\cite{Gurdogan:2015csr}. Therefore, the
theory~(\ref{eq:gam_def}) simplifies considerably and boils down
to~\cite{Gurdogan:2015csr,Kazakov:2018hrh}
\be
\label{eq:FCFT_ST}
\mathscr L = N _ c \,\Tr \l( \p _ \m \xoverline X \p _ \m X +\p_\m
\xoverline Z \p _ \m Z + \tilde\xi ^ 2 \xoverline X \xoverline Z X Z\r
) \ ,
\ee
where $\phi_1=X, \phi_2=Z$ and $\tilde\xi =4\pi\,\xi$.\footnote{The
factor of $4\pi$ was introduced for convenience.}~Notice that by
taking that specific limit, the Hermitian counterpart of the
interaction term does not survive,\footnote{To explicitly see how this
comes about, it is useful to expand the anticommutators in the
potential of~(\ref{eq:gam_def}); the terms of interest are
\be
\label{eq:intermediate}
\mathscr L \supset N _ c \,\Tr \l( g_{YM}^2e^{-i\g_3} \xoverline X
\xoverline Z X Z +g_{YM}^2e^{i\g_3} \xoverline Z \xoverline X Z X \r )
+\ldots \ .
\ee
In the DS limit~(\ref{eq:DS_limit}),~(\ref{eq:DS_limit_2}), it is
obvious that the coefficient of the last term decays
exponentially. This is the source of the non-Hermiticity of the
single-trace term.}~something that has far-reaching
implications. Namely, the number of planar diagrams at each order in
perturbation theory is severely restricted to the point that only a
handful of them needs to be computed (the number of course depends on
the quantity under consideration). Moreover, the multiloop diagrams of
the theory arrange themselves into regular square lattices with
quartic massless $\phi^4$-type vertices, so they resemble a
``fishnet.'' ~Such graphs are integrable~\cite{Zamolodchikov:1980mb},
see also~\cite{Chicherin:2012yn}. Owing to that, the FCFT is
integrable in the planar
limit~\cite{Gurdogan:2015csr,Caetano:2016ydc,Gromov:2017cja}, which
translates into nontrivial CFT data$=\{$Operator Product Expansion
(OPE) coefficients, scaling dimensions $\Delta\}$ be in principle
calculable~\cite{Gromov:2018hut}. As it will become clear in the next
sections, the non-Hermiticity of the biscalar FCFT gives rise to a
number of remarkable properties when its comes to conformal symmetry
breaking too, especially when compared to other nonsupersymmetric
CFTs.

It is important to note at this point that the theory as given
by~(\ref{eq:FCFT_ST}) is not complete, starting already from the
one-loop order. Rather, it should be supplemented by the following
double-trace terms~\cite{Fokken:2013aea}
\be
\label{eq:double-trace}
\begin{aligned}
\mathscr L _{d.t.}/(4\pi)^2 = \a _ 1 ^ 2 \l [ \Tr(X^2)\Tr(\xoverline X
^ 2) +\Tr(Z^2)\Tr(\xoverline Z ^ 2)\r ]-\a _ 2 ^ 2 \l [ \Tr
(XZ)\Tr(\xoverline X \xoverline Z) +\Tr(X\xoverline Z)\Tr(\xoverline X
Z) \r ]& \ ,
\end{aligned}
\ee
that are required in order to renormalize the correlators of the
composite operators $\Tr(X^2)$, $\Tr(\xoverline X^2)$, $\Tr(\xoverline
X Z)$, $\Tr( X \xoverline Z)$. In the above, $\a _ 1$ and $\a _ 2$ are
couplings whose running with the renormalization scale is responsible
for the explicit breaking of conformal symmetry. It turns out,
however, that the beta functions for the couplings have the following
two complex fixed points parametrized by
$\xi$~\cite{Sieg:2016vap,Grabner:2017pgm}
\be
\a_1 ^ 2=\a_+ ^ 2 \ ,~~~\a_2 = \xi ^2~~~~~~\text{and}~~~~~~\a_1 ^
2=\a_- ^ 2 \ ,~~~\a_2 = \xi ^2 \ ,
\ee
with
\be
\label{eq:a_1}
\a _ {\pm} ^ 2 =\pm \f {i\xi ^ 2 }{2} - \f {\xi ^
4 } {2} \mp \f {3i\xi ^ 6} {4} +\op (\xi ^ {8})\ . 
\ee 
When the couplings take their critical values, the FCFT
\be
 \mathscr L_{\rm FCFT} =  \mathscr L + \mathscr L_{d.t.} \ ,
\ee
behaves as a fully-fledged finite conformal theory for arbitrary
values of $\xi$.

\section{Classical vacua and their quantum fate}
\label{sec:flat_directions}

\subsection{Classical considerations}

The (matrix) equations of motion follow easily by varying the action
\be
S = \int d ^ 4 x~\mathcal L _ {\rm FCFT} \ ,  
\ee
w.r.t.~$X,\xoverline{X},Z,\xoverline{Z}$. For constant field
configurations, these respectively read
\be
\label{eq:eoms}
\begin{aligned}
&\kappa \,\Tr ( \xoverline X ^ 2 ) X +
\Tr ( \xoverline X Z) \xoverline Z+ \Tr (\xoverline X \xoverline Z) 
Z = N _ c Z \xoverline X \xoverline Z \ , \\
&\kappa \,\Tr ( X ^ 2 ) \xoverline X + \Tr ( X \xoverline Z) 
Z + \Tr ( X Z) \xoverline Z = N _ c \xoverline Z X Z  \ ,\\
&  \kappa \,\Tr(\xoverline Z ^ 2) Z+ \Tr ( X \xoverline Z)
\xoverline X+\Tr (\xoverline X \xoverline Z)  X = N _ c 
\xoverline X \xoverline  Z X \ ,\\
& \kappa\, \Tr ( Z ^ 2 ) \xoverline Z +\Tr ( \xoverline X Z) X + 
\Tr ( X Z) \xoverline X = N _ c X Z \xoverline X \ ,
\end{aligned}
\ee
where $\kappa = -2 \a^2_{\pm}/\x^2$. Notice that the equations for the
fields and their Hermitian transposes are not related by complex
conjugation. This follows from the particular form of the single-trace
term and the fact that $\kappa$ is complex. This is nothing but the
manifestation of the non-unitarity of the theory at the level of the
equations of motion.

The question to be addressed is if there exist vacua\,\footnote{Keep
in mind that the theory is not unitary, so in an abuse of language, by
``vacua'' and ``ground states'' I actually mean extrema of the complex
action.}~of the theory for which at least one of the field's vev is
nonvanishing, meaning that the symmetry is nonlinearly realized. To
proceed, let me make the simplest possible ansatz, i.e. require that
the vacuum expectation value (vev) of one of the fields is zero
\be
\label{eq:Xvev}
\av X _ {\rm tree} = 0 \ ,
\ee
while the other has a diagonal vev, 
\be
\label{eq:Zvev}
\av Z _ {\rm tree} = v\, \text{diag}\l (z_1, \ldots,z_{N_c} \r ) \ ,
\ee
with $v$ a (complex) parameter carrying mass dimension; the elements
$z _ k$ are complex numbers that satisfy
\be
\label{eq:constr_1}
\sum _ {k=1} ^ {N _ c} z _ k= \sum _ {k=1} ^ {N _ c}\bar z _ k= 0 \ ,
\ee
since the matrix fields are traceless. For obvious reasons, in what
follows I will refer to~(\ref{eq:Xvev}) and~(\ref{eq:Zvev}) as
``asymmetric vacua.''

One can immediately verify that the first two equations of motion are
identically satisfied on the above ansatz, while the last two yield
\be
\kappa\,\Tr\l(\av{\xoverline Z} _ {\rm tree}^ 2\r) \av Z _ {\rm tree}=
0,~~\text{and}~~\kappa\,\Tr\l(\av Z _ {\rm tree} ^2\r) \av{\xoverline
  Z} _ {\rm tree}= 0 \ ,
\ee
which translates into 
\be
\label{eq:constr_2}
\sum _ {k=1} ^ {N_c} z _ k ^ 2 = \sum _ {k=1} ^ {N_c} \bar z _ k ^ 2=
0 \ ,
\ee
since both $\kappa$ and $\av Z _{\rm tree}$ are non-zero. 

It is impressive that the theory supports flat directions with
vanishing energy for every value of the coupling. Put differently,
there is actually no need for finetuning in order for the symmetry
breaking
vacua~(\ref{eq:Zvev}),~(\ref{eq:constr_1}),~(\ref{eq:constr_2}) to
become accessible to the system.  Although this is a rather unusual
situation for a CFT without supersymmetry---flat directions were
believed to appear only for particular values of the corresponding
couplings~\cite{Fubini:1976jm,Shaposhnikov:2008xb,Coradeschi:2013gda}---it
is not a mystery. The $\mathcal N = 4$~SYM has a plethora of vacua
that nonlinearly realize the conformal symmetry. The FCFT, in spite of
being a heavily deformed descendant of this theory, has nevertheless
inherited some of the aforementioned flat directions.

That's not the end of the story though. Provided that $\av X _ {\rm
tree} =0$ there are more symmetry breaking solutions
to~(\ref{eq:eoms}), also for all values of {$\x$}. These are absent
both in the $\mathcal N=4$~SYM, as well as its
$\gamma$-deformation and appear only after the DS Fishnet
limit~(\ref{eq:DS_limit}),~(\ref{eq:DS_limit_2}) is taken. Such
solutions are nilpotent matrices for which $\av Z _ {\rm tree}\neq 0$,
with $\av Z _ {\rm tree} ^ 2 =0$. Note that additional flat directions
appear $\forall \xi$ if both fields acquire a vev, for instance $\av X
_ {\rm tree} \propto \av Z _ {\rm tree}$, with $\av Z _ {\rm tree}$
given by~(\ref{eq:Zvev}), as well as for specific values of the
coupling. It is surely worth investigating the implications of having
such a rich moduli space, nevertheless in what follows I will only
focus on the simplest, asymmetric vacua.

\subsection{Quantum corrections}

Although conformal symmetry can certainly be present at the classical
level, it may be explicitly broken when quantum corrections are taken
into account. This would be a rather problematic situation, since the
classical flat directions are lifted and the vacuum energy does not
vanish anymore. To put it in other words, the dilaton acquires mass.

To investigate the quantum fate of the classical symmetry-breaking
vacua, one needs to compute the Coleman-Weinberg (CW) effective
potential~\cite{Coleman:1973jx}. Here, I will confine myself to the
one-loop level, since a possible uplifting of the flat vacua is
visible already at this order.

The computation proceeds as follows. First, it is convenient to
rescale the fields by $\sqrt N_c$ to make their kinetic terms
canonical. To have the correct counting of $N_c$, the double trace
terms should be factorized by the introduction of auxiliary Lagrange
multiplier fields $A,B,\ldots$~\cite{Coleman:1974jh}; the
non-derivative part of the theory thus becomes
\be
\label{eq:potential_lagr_mult}
\begin{aligned}
N _ c V/ \tilde\xi^2&= \Tr\l ( \xoverline X \xoverline Z X Z \r ) -
\xoverline A\,\Tr (X^2) - A\, \Tr(\xoverline X ^2 ) - \xoverline
B\,\Tr (Z^2) - B\, \Tr(\xoverline Z ^2 ) -\xoverline C\,\Tr (X Z) -
C\, \Tr(\xoverline X \xoverline Z )\\ & -\xoverline D\,\Tr (X
\xoverline Z) - D\, \Tr(\xoverline X Z )+ \f {2N_c}{\kappa} \xoverline
A A + \f {2N_c}{\kappa} \xoverline B B +N _ c\xoverline C C + N _ c
\xoverline D D\ .
\end{aligned}
\ee
Then, one expands the fields around a vacuum configuration 
\be
X =
\av X + \d X,~Z = \av Z + \d Z \ ,
\ee
where $\av X$ and $\av Z $ are arbitrary constant $SU(N_c)$
matrices. The effective potential reads
\be
\label{eq:Veff}
V_{\rm eff} = V + V _{\rm 1-loop} \ ,
\ee
where the one-loop
correction is schematically given by
\be
V _ {\rm 1-loop}\propto\Tr\left( M^4 \log \frac{M^2}{\m^2} \right)  \ ,
\ee
with $M$ the ``mass matrix;'' $\m$ is the 't Hooft-Veltman
renormalization point.

\begin{figure}[!t]
\centering
\includegraphics[scale=.35]{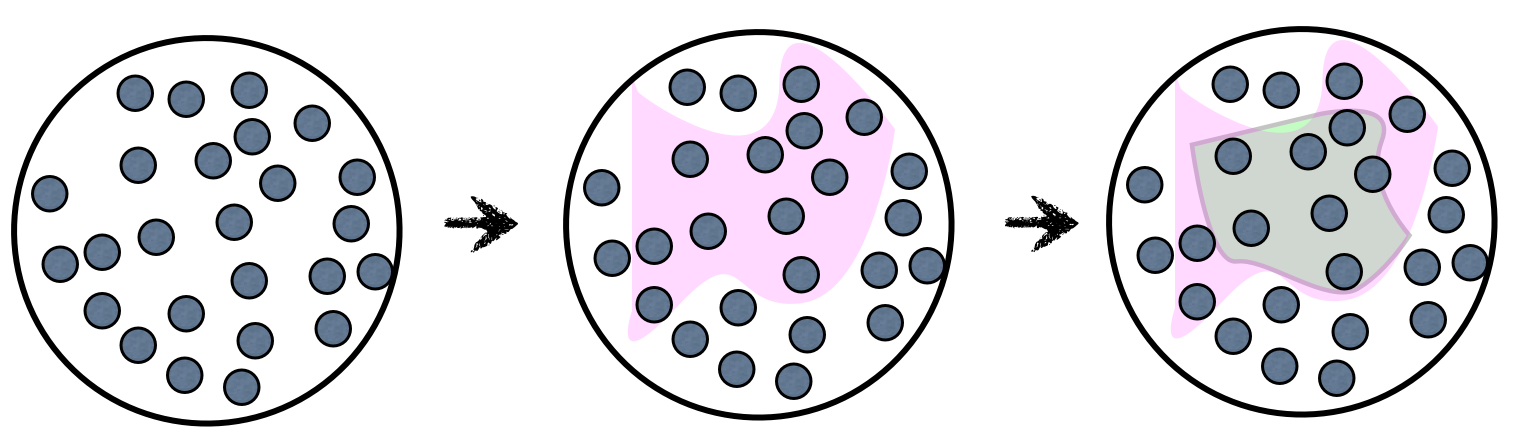}
\caption{The parent $\mathcal N=4$ supersymmetric Yang-Mills theory
possesses a plethora of nontrivial flat directions. Some of them are
passed down to the FCFT (pink color). Most importantly, a nonzero set
of them is not lifted by quantum corrections (green color). On top of
these symmetry-breaking vacua, the vacuum energy of the theory is
naturally zero.}
\label{fig:vacua}
\end{figure}

The extrema of~(\ref{eq:Veff}) are determined by requiring that the
first derivatives of the potential w.r.t.~all the fields vanish. A
detailed computation~\cite{Karananas:2019fox} reveals that a subclass
of the asymmetric classical
vacua~(\ref{eq:Zvev}),~(\ref{eq:constr_1}),~(\ref{eq:constr_2}) is
robust under quantum corrections. This corresponds to setting
\(A=B=C=D=0\), and imposing the following extra constraints on the
matrix elements
\begin{equation}
\label{eq:constr_3}
\sum _ {k=1} ^ {N _ c} z _ k ^ 2 \log z _ k = \sum _ {k=1} ^ {N _ c}
\bar z _ k ^ 2 \log \bar z _ k  = 0 \ .
\end{equation}
The vacuum energy of the quantum corrected theory in this Coulomb
branch is~\emph{naturally}~zero,
\be
V_{\rm eff} = 0 \ ,
\ee
or in other words, the dilaton continues being massless at the
one-loop level, without finetunings.

It was already pointed out that the FCFT has inherited many of the
vacua of its parent $\mathcal N=4$~SYM. Although by no means
guaranteed, out of them there is a nontrivial subclass which is
singled out since they survive quantum corrections, see also
Fig.~\ref{fig:vacua}.

\begin{figure}[!t]
\centering
\includegraphics[scale=.4]{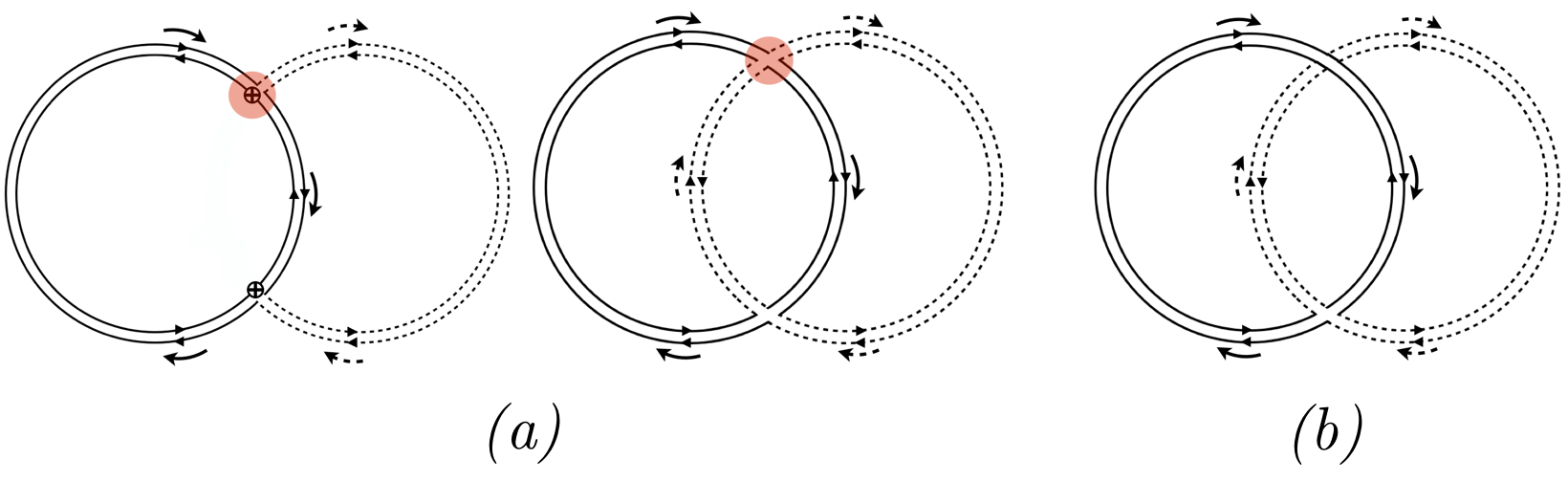}
\caption{\emph{(a)}~Diagrams at order $\tilde\xi^4$ that would in
principle contribute to the CW potential. A solid (dashed) line stands
for the excitations of $X$ ($Z$), and ``$\pmb{\otimes}$'' for the vev
of $Z$. The absent antichiral vertices have been marked with
red.~\emph{(b)}~A non-planar diagram with one cubic and one quartic
vertex that should not be taken into account in the evaluation of the
effective potential. }
\label{fig:forbidden_1}
\end{figure}

Before moving on, it is worth pointing out yet another nontrivial
feature of the FCFT: many higher loop diagrams that can in principle
spoil the flatness of the effective potential, are absent in the
planar limit. This is due to the constraints
~(\ref{eq:constr_1}),~(\ref{eq:constr_2}),~(\ref{eq:constr_3}) that
force them to vanish on the top of the vacua under consideration. In
addition, the non-Hermiticity/fixed chirality of the single-trace
interaction acts as a self-protection mechanism in the sense that the
``antichiral'' vertices that would stem from the nonexistent
$\Tr(\xoverline Z\xoverline X Z X)$ term are now absent. Consequently,
another class of dangerous diagrams, such as those presented in
Fig.~~\ref{fig:forbidden_1}\emph{(a)}, cannot be constructed at
all. It remains to be understood whether the theory receives no
multi-loop contributions at all.

Note, however, that the flatness of the effective potential may be
spoiled due to the presence of diagrams like the one of
Fig.~\ref{fig:forbidden_1}\emph{(b)}, which appears at the $1/N_c^2$
order. In general, it would not be surprising that some properties
of the FCFT do not survive beyond the planar limit (or for finite
number of colors), even when the conformal symmetry is linearly
realized. This is certainly something that requires separate
investigations.

\subsection{Quantum vacua: an existence proof}

It is instructive to actually present a simple example of a vacuum
which satisfies the aforementioned constraints, so the flatness of
potential is not ruined. Consider $\av X=0$, while the expectation
value of the other field $\av Z$ is an $N_c\times N_c$ block-diagonal
matrix comprising $N _ c/4$ diagonal sub-blocks each with dimensions
$4\times 4$, i.e.
\begin{equation}
\label{eq:block-mat-Z}
\av Z = v 
  \begin{tikzpicture}[baseline=-0.5ex]
  \footnotesize{
\matrix [matrix of math nodes,left delimiter=(,right delimiter=)] 
   { z_1 & & & & & & & & \\
     & z_2 & & & & & & &  \\
      & & z _3 & & & & & &  \\
      & & & z _ 4  && & & &  \\
     & & & & \ddots & & & &  \\
     & & & & &  z _ 1  & & &  \\
     & & & & & & z _ 2 & &  \\
     & & & & & & & z _ 3  & \\
     & & & & & & & & z _ 4 \\
   };}
   \draw (-2.2,2.0) rectangle (-0.3,0.3);
   \draw (2.2,-2.0) rectangle (0.3,-0.3);
 \end{tikzpicture} \ .
\end{equation}
With this ansatz, the set of the transcendental
equations~(\ref{eq:constr_1}),~(\ref{eq:constr_2}),~(\ref{eq:constr_3})
admits the following complex numerical solution
\be
\begin{aligned}
\label{eq:explicit_su4}
&z_1 = -0.587849-0.808971\,i\ ,~~~z_2 =0.260305+  1.45187\,i
      \ ,\\ &z_3 =1.32754- 0.642903\,i\ ,~~~~~~\,\,\,\,\, 
z_4 =-1 \ .
\end{aligned}
\ee

More (complicated) vacua can be constructed in a straightforward
manner. For instance, one can seek for solutions to
eqs.~(\ref{eq:constr_1}),~(\ref{eq:constr_2}),~(\ref{eq:constr_3}) by
considering bigger sub-blocks of the same dimension, or even
combinations of sub-blocks of different dimensionalities.

\section{Conclusions}
\label{sec:conclusions}

In the present talk I touched upon conformal symmetry breaking in the
context of the FCFT. As it hopefully became clear, this is a rather
special theory for a variety of reasons.

First, the theory can accommodate nontrivial vacua without the burden
of finetuning the corresponding coupling(s). This is in one-to-one
with the dynamical generation of scales, while the vacuum energy is
zero, naturally.

Importantly, in the planar limit, the subclass of the classical
asymmetric flat directions $\av X = 0,~\av Z\neq 0$ given
by~(\ref{eq:Zvev}) and subject
to~(\ref{eq:constr_1}),~(\ref{eq:constr_2}) and~(\ref{eq:constr_3}),
was shown to not be lifted by quantum effects, at least at the
one-loop order. In other words, the FCFT exhibits nonlinear
realization of exact quantum conformal symmetry.

Such unique features are the aftermath of:~\emph{i)}~the model's
non-Hermiticity that translates into a fixed orientation of the
interaction vertices; ~\emph{ii)}~the fact that the considerations
concern the large-$N_c$
limit;~\emph{iii)}~the~constraints~(\ref{eq:constr_1}),~(\ref{eq:constr_2})
and~(\ref{eq:constr_3}) on the flat directions.

Although the phenomenological relevance of the FCFT per se is not
clear, the theory is an ideal playground for studying the dynamics
underlying the spontaneous breaking of quantum conformal symmetry. For
example, confronting it with the consistency conditions on the CFT
data presented in~\cite{Karananas:2017zrg}, may shed light on general
properties shared also by realistic effective field theories (see also
footnote~\ref{foot:refs}) exhibiting nonlinearly realized exact
conformal invariance. This can be a step towards understanding to what
extent CFTs are in the heart of the solutions to the Standard Model
finetuning issues.

\acknowledgments{I am grateful to V.~Kazakov and M.~Shaposhnikov for
the collaboration and for numerous discussions. It is a pleasure to
thank the organizers of the ``Conference on Recent Developments in
Strings and Gravity'' for the invitation and the stimulating
atmosphere. This work was partially supported by the Deutsche
Forschungsgemeinschaft (DFG, German Research Foundation) under
Germany’s Excellence Strategy EXC–2111–390814868.}

\bibliographystyle{JHEP}
{\small\setlength{\bibsep}{4.5pt plus 1ex}
\bibliography{Broken_CFT}}
\end{document}